\documentclass[aps,twocolumn,groupedaddress,showpacs]{revtex4}
\usepackage{graphicx}
\begin{document}
\bibliographystyle{apsrev}


\title{Observation of Bulk Superconductivity in Na$_x$CoO$_2\cdot$yH$_2$O and Na$_x$CoO$_2\cdot$yD$_2$O Powder and Single Crystals}


\author{R. Jin$^1$}
\email[]{email address: jinr@ornl.gov}
\author{B.C. Sales$^1$}
\author{P. Khalifah$^{1}$}
\author{D. Mandrus$^{1,2}$}

\affiliation{$^1$Condensed Matter Sciences Division, Oak Ridge
National Laboratory, Oak Ridge, TN 37831}
\affiliation{$^2$Department of Physics and Astronomy, The
University of Tennessee, Knoxville, TN 37996}


\date{\today}

\begin{abstract}
Poly- and single-crystalline Na$_x$CoO$_2$ has been successfully
intercalated with H$_2$O and D$_2$O as confirmed by X-ray
diffraction and thermogravimetric analysis. Resistivity, magnetic
susceptibility, and specific heat measurements show bulk
superconductivity with T$_c$ close to 5 K in both cases. The
substitution of deuterium for hydrogen has an effect on T$_c$ of
less than 0.2 K. Investigation of the resistivity anisotropy of
Na$_x$CoO$_2\cdot$yH$_2$O single crystals shows: (a) almost zero
resistivity below T$_c$, and (b) an abrupt upturn at $T^* \sim$ 52
K in both the $\it {ab}$ plane and the $\it {c}$ direction. The
implications of our results on the possible superconducting
mechanism will be discussed.

\end{abstract}

\maketitle

Sodium-cobalt oxide Na$_x$CoO$_2$ has been of interest for several
years as a potential thermoelectric material, because it exhibits
low resistivity coupled with a relatively large thermopower
\cite{Terasaki}. The crystal structure consists of layers of
edge-sharing CoO$_6$ octahedra perpendicular to the $\it{c}$ axis
separated by Na layers.  The Co ions are mixed-valent with a
formal oxidation state of 4 - $\it {x}$.  Qualitatively, this
structure is similar to that of high-T$_c$ cuprate
superconductors, except that in each layer the Co atoms form a
triangular (hexagonal) lattice rather than a square lattice. The
DC magnetic susceptibility $\chi$ of Na$_x$CoO$_2$ is small and
weakly temperature dependent, with no evidence of long-range
magnetic order for $\it {x}$ $<$ 0.75.  For $\it {x}$ = 0.75, a
weak magnetic transition at 21 K was reported by Motohashi and
co-workers \cite{Motohashi}. Magnetic frustration within the Co
layers likely suppresses robust long-range magnetic order. The
magnetic susceptibility and the in-plane ($\rho_{ab}$) and
out-of-plane ($\rho_c$) resistivities of Na$_{0.5}$CoO$_2$ are, in
fact, reminiscent of another layered superconductor, Sr$_2$RuO$_4$
\cite{Terasaki,Maeno}, except for the absence of superconductivity
at low temperatures. The recent discovery of superconductivity
\cite{Takada} in water-intercalated
Na$_{0.35}$CoO$_2\cdot$1.3H$_2$O is very exciting, because it may
be the first superconductor analogous to Sr$_2$RuO$_4$ as proposed
by Tanaka and Hu \cite{Tanaka}.

Although $\rho_{ab}$ is small, superconductivity has not been
observed in Na$_x$CoO$_2$. Superconductivity also does not occur
in Na$_{0.3}$CoO$_2\cdot$0.6H$_2$O \cite{Foo}, indicating that
sufficient hydration is crucial for the appearance of
superconductivity in Na$_{0.3}$CoO$_2\cdot$yH$_2$O. The
relationship between the transition temperature T$_c$ and the
water content $\it {y}$, and the role that H$_2$O plays for the
occurrence of superconductivity are central issues. In this
Letter, we report that (a) the results reported in Ref.
\cite{Takada} can be reproduced, (b) the superconducting volume
may be enhanced or suppressed, depending on the water content, (c)
superconductivity is observed in deuterated
Na$_{0.3}$CoO$_2\cdot$1.4D$_2$O with T$_c \sim$ 4.5 K, and (d) the
superconductivity is observed in both the $\it {ab}$-plane and the
$\it {c}$-axis resistivities ($\rho_{ab}$ and $\rho_c$) of
Na$_x$CoO$_2\cdot$yH$_2$O single crystals.

 Superconducting sodium cobaltate was prepared following a procedure similar to that
  described in Ref. \cite{Takada}.  The parent compound Na$_x$CoO$_2$ was
 obtained via solid-state reaction.  Starting materials Na$_2$CO$_3$ (Alfa Aesar 99.997\%)
 and Co$_3$O$_4$ (Alfa Aesar 99.9985\%)
 were mixed in a molar ratio of Na:Co = 0.7:1.0.  After ball milling in a sealed chamber for 2 h,
 the mixture was put into a furnace that was preheated to 750 $^\circ$C, $\it
 {i.e.}$, the so-called
  rapid-heat-up technique \cite{Motohashi}.  After heating for 20 h,
  the powder was re-ground, pressed into pellets, and calcined at 830 $^\circ$C for 16 h
  in flowing O$_2$ gas.  X-ray diffraction results confirm the correct single phase with
  hexagonal crystal structure (lattice parameters are listed in Tab.\ 1).  For single
  crystal growth, both the floating-zone technique
  and flux method \cite{Fujita} were employed.  In the latter case, thin
plate-like single crystals were obtained
  with surface areas up to 5$\times$5 mm$^2$.  Even larger single crystals could be grown using
 the floating-zone technique.  Based on the values of the lattice parameters
   and high-temperature resistivity measurements, the actual Na content $\it {x}$ is close to 0.6
    for both poly- and
   single- crystals \cite{Motohashi,Foo}.  It is worth mentioning
   that, in addition to forming Na$_x$CoO$_2$, the
    conventional slow-heat-up process often results in impurity phases including Co$_3$O$_4$ and
    perhaps CoCO$_3$.

To obtain superconductivity, it is necessary to first chemically
extract additional Na from the structure. Both Na$_{0.6}$CoO$_2$
powder and small single crystals (1$\times$1$\times$0.1 mm$^3$)
were placed in a 6.6 molar Br$_2$/CH$_3$CN solution for 2-5 days.
Careful filtering and washing in pure CH$_3$CN followed by a 50/50
mixture of CH$_3$CN/H$_2$O (or CH$_3$CN/D$_2$O) resulted in a
single phase, intermediate state of hydration/deuteration first
reported by Foo $\it {et}$ $\it {al.}$ \cite{Foo}.  Our TGA
measurements on this phase give an approximate Na content $\it
{x}$ = 0.3 and H$_2$O/D$_2$O content of $\it {y}$ = 0.9.  This
value of $\it {y}$ is larger than $\it {y}$ = 0.6 reported in Ref.
\cite{Foo}. This phase is stable as long as the powder is kept in
a sealed bottle. As listed in Tab.\ 1, the lattice parameter $\it
{a}$ remains virtually unchanged, while the $\it {c}$ value is
enlarged to 13.831 $\AA$, a 26$\%$ increase compared to the parent
compound. Superconductivity was obtained through further
hydration/deuteration of the samples by stirring the powder or
small crystals in distilled H$_2$O or D$_2$O for more than 12 h at
room temperature. The exact time required to reach the optimum
state of hydration/deuteration ($\it {y}$ $\sim$ 1.4) for
superconductivity depends on the crystallite size. It is also
possible to overhydrate the powder ($\it {y}$ $\sim$ 1.8-2.0)
which strongly suppresses superconductivity but has little effect
on the X-ray pattern. As may be seen in Tab.\ 1, the lattice
parameters for Na$_{0.3}$CoO$_2\cdot$1.4H$_2$O are slightly larger
than that for Na$_{0.3}$CoO$_2\cdot$1.4D$_2$O, consistent with the
stronger D - O bond relative to H - O bond \cite{Shriver}.

\begin{figure}
\includegraphics[keepaspectratio=true, totalheight =3.5 in, width =
3.0 in]{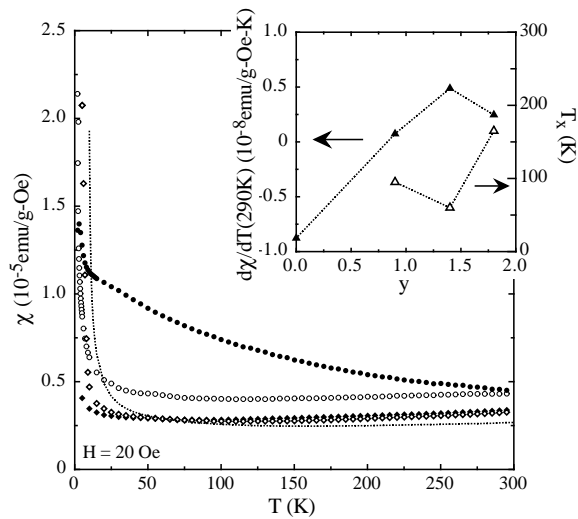} \caption{Temperature dependence of the DC
magnetic susceptibility for the parent compound Na$_{0.6}$CoO$_2$
(filled circles), intermediate phase
Na$_{0.3}$CoO$_2\cdot$0.9H$_2$O (unfilled circles),
superconducting Na$_{0.3}$CoO$_2\cdot$1.4H$_2$O (filled diamonds)
and Na$_{0.3}$CoO$_2\cdot$1.4D$_2$O (unfilled diamonds) above the
superconducting transition T$_c$, and overdeuterated
Na$_{0.3}$CoO$_2\cdot$1.8D$_2$O (dashed line).  All measurements
were carried out by applying H = 20 Oe using a commercial SQUID
magnetometer by Quantum Design. The inset shows the H$_2$O/D$_2$O
content $\it {y}$ dependence of the room-temperature magnetic
susceptibility slope d$\chi$/dT(290K) (filled triangles) and
crossover temperature T$_x$ (unfilled triangles) (see the
definition in the text). Note that, both d$\chi$/dT(290K) and
T$_x$ exhibit extrema at $\it {y}$ = 1.4, indicating that the
susceptibility data correlates with both the Na content (Co
valence) and the degree of hydration/deuteration.}
\end{figure}

In Fig.\ 1, we present the temperature dependence of the DC
magnetic susceptibility $\chi$ of Na$_{0.6}$CoO$_2$,
Na$_{0.3}$CoO$_2\cdot$yH$_2$O ($\it {y}$ = 0.9, 1.4),
 and Na$_{0.3}$CoO$_2\cdot$yD$_2$O ($\it {y}$ = 1.4, 1.8). Note that for the
parent compound $\chi$ is positive and increases with decreasing
temperature. Although the Curie-Weiss-like tail remains below a
particular temperature T$_x$, the oxidation and
hydration/deutration process results in qualitative changes in the
magnetic susceptibility of Na$_{0.3}$CoO$_2\cdot$yH$_2$O and
Na$_{0.3}$CoO$_2\cdot$yD$_2$O at high temperatures.  Above T$_x$,
$\chi$ increases with T, in contrast to that of the parent
compound. As shown in the inset of Fig.\ 1, the smaller T$_x$, the
larger the slope d$\chi$/dT determined at T = 290 K. However, both
T$_x$ and d$\chi$/dT(290K) vary non-monotonically with $\it {y}$,
exhibiting extrema at $\it {y}$ = 1.4.  This demonstrates that the
qualitative change in $\chi$ depends on the sodium content (cobalt
valence) as well as the degree of hydration/deuteration. For fixed
$\it {y}$, the substitution of deuterium for hydrogen results in
little effect on the high-temperature magnetic susceptibility.

\begin{table}
\includegraphics[keepaspectratio=true, totalheight = 2.0 in, width = 3.0 in]{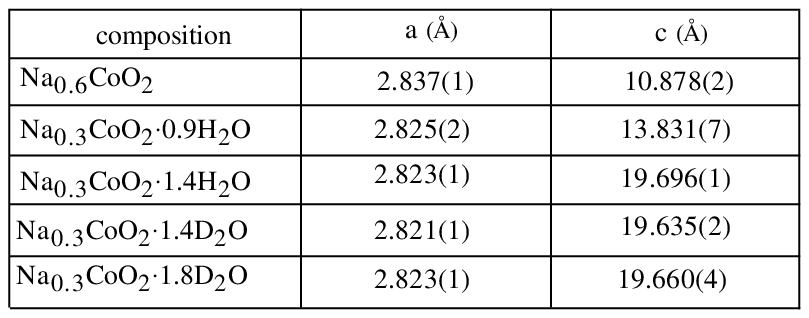}
\caption{Room temperature lattice parameters and standard
deviations for Na$_{x}$CoO$_2\cdot$yH$_2$O and
Na$_{x}$CoO$_2\cdot$yD$_2$O (space group no. 194 P6$_3$mmc).}
\end{table}

As mentioned previously, superconductivity does not occur until
the water content $\it {y}$ is close to 1.4. Presented in Fig.\ 2
is the temperature dependence of the magnetic susceptibility and
resistivity between 1.9 and 10 K for polycrystalline
Na$_{0.3}$CoO$_2\cdot$1.4H$_2$O and
Na$_{0.3}$CoO$_2\cdot$1.4D$_2$O. Under both zero-field-cooling
(zfc) and field-cooling (fc) conditions, $\chi$ becomes negative
below T$_c$ = 4.5 K in both materials.  Correspondingly, the
electrical resistivity measured on pressed pellets departs from
high-temperature behavior and drops rapidly below T$_c$ (see Fig.\
2c-2d), indicating that the system undergoes a superconducting
transition. The substitution of deuterium for hydrogen has no
apparent effect on T$_c$. While $\chi_{zfc}$ is much larger than
that reported previously \cite{Takada,Foo}, $\chi_{fc}$ is still
small for both Na$_{0.3}$CoO$_2\cdot$1.4H$_2$O and
Na$_{0.3}$CoO$_2\cdot$1.4D$_2$O.  Note that the electrical
resistivity of the polycrystalline samples does not vanish down to
1.9 K, similar to the results reported in Ref. \cite{Takada}.
These observations imply that the superconducting volume fraction
may be improved.

\begin{figure}
\includegraphics[keepaspectratio=true, totalheight = 3.5 in, width =
3.2 in]{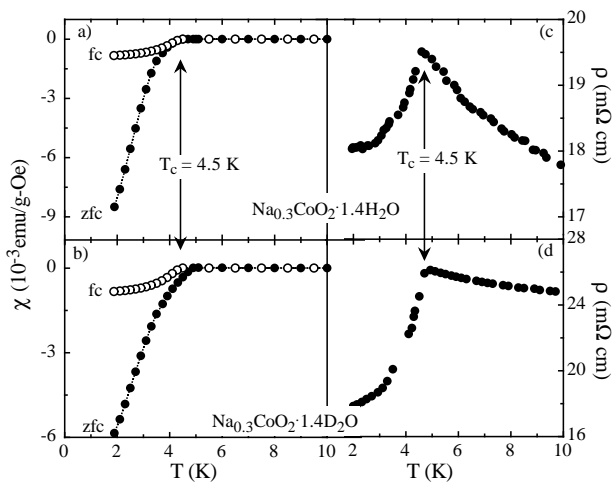} \caption{Temperature dependence of the
magnetic susceptibility $\chi$ of Na$_{0.3}$CoO$_2\cdot$1.4H$_2$O
(a) and Na$_{0.3}$CoO$_2\cdot$1.4D$_2$O (b) between 1.9 and 10 K
at H = 20 Oe.  The measurements were performed under both
zero-field-cooling (zfc) (filled circles) and field-cooling (fc)
(unfilled circles) conditions. Shown in right panels are the
temperature dependences of the electrical resistivity of
polycrystalline Na$_{0.3}$CoO$_2\cdot$1.4H$_2$O (c) and
Na$_{0.3}$CoO$_2\cdot$1.4D$_2$O (d). }
\end{figure}

Specific heat C$_p$ was measured on pellets prepared by pressing
the powder with the best superconducting volume fraction as
determined by magnetic susceptibility. In the inset (a) of Fig.\
3, we present C$_p$ versus T between 2 and 60 K. No obvious
anomaly is visible. However, the superconducting anomaly can be
clearly seen when the data is replotted as C$_p$/T versus T$^2$ as
shown in the main panel of Fig.\ 3. Note C$_p$/T varies more or
less linearly with T$^2$ above $\sim$ 7.5 K, suggesting that the
electron and phonon contributions dominate C$_p$ in this region.
The normal electronic specific heat coefficient may thus be
estimated via the linear fitting procedure. We obtain $\gamma
\sim$ 15.8 mJ/mol-K$^2$ for Na$_{0.3}$CoO$_2\cdot$1.4D$_2$O and
14.4 mJ/mol-K$^2$ for Na$_{0.3}$CoO$_2\cdot$1.4H$_2$O, close to
that reported in Ref. \cite{Cao}. In the lower temperature region,
C$_p$/T deviates from linearity in both systems, however. By
subtracting the phonon contribution (T$^2$ term in C$_p$/T), one
may clearly see that the electronic specific heat C$_p^{el}$
exhibits a hump centered around 6.0 K as displayed in the inset
(b) of Fig.\ 3. In combination with the magnetic susceptibility
and resistivity data, we believe that the hump in C$_p^{el}$/T
results from the superconducting transition. The lack of a sharp
specific heat peak may be due to the distribution of T$_c$ from
H$_2$O/D$_2$O deintercalation during pellet pressing process.

\begin{figure}
\includegraphics[keepaspectratio=true, totalheight = 3.5 in, width =
3.0 in]{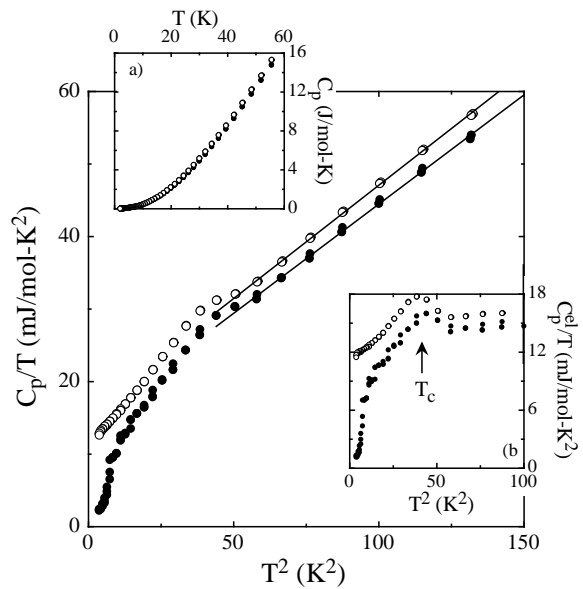} \caption{Temperature dependence of
specific heat C$_p$ of Na$_{0.3}$CoO$_2\cdot$1.4H$_2$O (filled
circles) and Na$_{0.3}$CoO$_2\cdot$1.4D$_2$O (unfilled circles)
plotted as C$_p$/T versus T$^2$ between 1.9 and 12 K. The solid
lines are the fit to experimental data between 7.5 and 12 K using
C$_p$/T = $\gamma$ + $\beta$T$^2$ ($\beta$ and $\gamma$ are
constants). The inset shows the temperature dependence of (a)
C$_p$ between 1.9 and 60 K and (b) electronic specific heat
C$_p^{el}$ plotted as C$_p^{el}$/T versus T$^2$ between 1.9 and 12
K. }
\end{figure}

We recall that there is no specific heat anomaly at T$_c$ in the
optimally doped Nd$_{1.85}$Ce$_{0.15}$CuO$_4$ superconductor,
while its electrical resistivity reaches zero just below T$_c$
\cite{Takagi}. So far, zero-resistivity has not be reported in
superconducting sodium-cobalt oxide. Similar to what is shown in
Ref. \cite{Takada}, the resistivity of our best polycrystalline
Na$_{0.3}$CoO$_2\cdot$1.4H$_2$O tends to saturate after 10 - 50\%
reduction compared to the normal-state value (see Fig.\ 2c-2d). Is
this intrinsic? From the structural point of view, the present
system may have very weak coupling between adjacent CoO$_2$
layers, as they are separated by two layers of H$_2$O/D$_2$O and
one layer of Na. If the superconductivity were confined within the
CoO$_2$ layers, the zero-resistivity state may never be observed
in polycrystals, due to the contribution of non-zero $\rho_{c}$.

Bearing this issue in mind, we have investigated the resistivity
anisotropy of superconducting Na$_{0.3}$CoO$_2\cdot$yH$_2$O single
crystals. To reduce the loss of H$_2$O, a silver paint that dries
at room temperature was used to adhere four leads onto each
crystal. The contact resistance was about 5 - 20 $\Omega$, after
drying at room temperature for 30 minutes.  Shown in Fig.\ 4 is
the temperature dependence of the electrical resistivity of
Na$_{0.3}$CoO$_2\cdot$yH$_2$O in the $\it {ab}$ plane (a) and
along the $\it {c}$ direction (b). Compared to the parent
compound, both $\rho_{ab}$ and $\rho_c$ are somewhat large at room
temperature. We believe this is due to the error involved in
estimating the geometric factor from small superconducting single
crystals with a
 typical size of 1$\times$1$\times$0.1 mm$^3$. Nevertheless, both
$\rho_{ab}$ and $\rho_c$ exhibit very similar behavior as that of
the parent compound above T$^*$ = 52 K.  While $\rho_{ab}$
decreases with T, $\rho_c$ reveals a broad maximum near 200 K as
indicated in Fig.\ 4b. Strikingly, below T$^*$, both $\rho_{ab}$
and $\rho_c$ increase with decreasing T before entering the
superconducting state.  Although the resistivity of
polycrystalline Na$_{0.35}$CoO$_2\cdot$1.3H$_2$O also increases
with decreasing T below T$^*$ (see Fig.\ 4 in Ref. \cite{Takada}),
the change is much smoother and less pronounced.  It should be
emphasized that such a sharp upturn at T$^*$ can only be observed
in single crystals, where $\rho_{ab}$ and $\rho_c$ exhibit an
abrupt decrease at T$_c$. This strongly suggests that the
resistivity upturn at T$^*$ is an intrinsic property of the
superconducting phase.  At present, it is unclear whether this new
feature is associated with a phase transition, since neither
magnetic susceptibility nor the specific heat data obtained on
polycrystalline samples show an anomaly in this temperature range.

\begin{figure}
\includegraphics[keepaspectratio=true, totalheight = 3.5 in, width =
3.5 in]{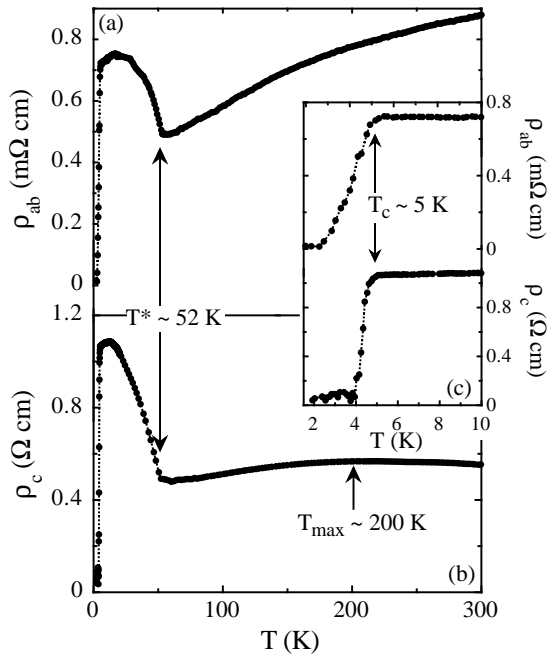} \caption{Temperature dependence of the
in-plane and out-of-plane resistivity $\rho_{ab}$ (a) and $\rho_c$
(b) of Na$_{0.3}$CoO$_2\cdot$yH$_2$O single crystals.  The inset
(c) is an enlargement of the low-temperature data, showing a
superconducting transition in both $\rho_{ab}$ and $\rho_c$ at
T$_c$ = 5 K. }
\end{figure}

Owing to the inaccurate geometric factor, the absolute value of
$\rho_{ab}$ and $\rho_c$ should be further examined.  However, it
is unambiguous that both $\rho_{ab}$ and $\rho_c$ drop
spontaneously below T$_c \sim$ 5 K.  This indicates
three-dimensional (3D) superconductivity, despite an extremely
high anisotropy ($\rho_{c}$/$\rho_{ab} \sim$ 10$^3$).
Surprisingly, the transition is sharper along the $\it {c}$ axis
than in the $\it {ab}$ plane, though at low temperatures both
$\rho_{ab}$ and $\rho_c$ saturate with a small but non-zero value
(see Fig.\ 4c).

Based on the above results, we believe that the coupling mechanism
between the CoO$_2$ layers is the key to understanding
superconductivity in this unique system. As shown in Fig.\ 1, the
magnitude of magnetic susceptibility tends to decrease with
increasing H$_2$O/D$_2$O content at room temperature. Most
prominent is that d$\chi$/dT(290K) exhibits a maximum at $\it {y}$
= 1.4, a value that also results in the highest T$_c$.
Correspondingly, the low-temperature tail is almost suppressed as
reflected by the smallest T$_x$. This indicates an intimate
relationship between the normal-state magnetism and
superconductivity as proposed by Tanaka and Hu \cite{Tanaka}.


\begin{acknowledgments}
We would like to thank J. He, Q. Kou, W. Tian and K. Affholter for
technical assistance. Oak Ridge National laboratory is managed by
UT-Battelle, LLC, for the U.S. Department of Energy under contract
DE-AC05-00OR22725.
\end{acknowledgments}

\bibliography{NCOH2Obib.tex}

%
%

%
%

\end{document}